\documentstyle[aps,prb,multicol,epsfig,amsmath]{revtex}
\renewcommand{\narrowtext}{\begin{multicols}{2}\global\columnwidth20.5pc}
\renewcommand{\widetext}{\end{multicols}\global\columnwidth42.5pc}

\newcommand{\wideequationend}{
\hfill
\begin{picture}(3.375,0)
  \put(0,0){\line(1,0){3.375}}
  \put(0,0){\line(0,-1){0.08}}
\end{picture}
\narrowtext}

\begin{document}

\draft
\title{Mesoscopic fluctuations of Coulomb drag between quasi-ballistic 1D--wires}

\author{Niels Asger Mortensen}

\address{Mikroelektronik Centret, Technical University of Denmark,
  {\O}rsteds Plads, Bld. 345 east, DK-2800 Kgs. Lyngby, Denmark\\ {\O}rsted Laboratory,
  Niels Bohr Institute for APG,
  Universitetsparken 5, DK-2100 Copenhagen {\O}, Denmark\\
Instituut-Lorentz, Universiteit Leiden, P.O. Box 9506, 2300 RA Leiden, The Netherlands}

\author{Karsten Flensberg}

\address{{\O}rsted Laboratory, Niels Bohr Institute for APG,
  Universitetsparken 5, DK-2100 Copenhagen {\O}, Denmark}

\author{Antti-Pekka Jauho}

\address{Mikroelektronik Centret, Technical University of Denmark,
  {\O}rsteds Plads, Bld. 345 east, DK-2800 Kgs. Lyngby, Denmark}

\date{\today}
\maketitle

\begin{abstract}
  Quasiballistic 1D quantum wires are known to have a conductance of
  the order of $2e^2/h$, with small sample-to-sample fluctuations. We
  present a study of the transconductance $G_{12}$ of two
  Coulomb-coupled quasiballistic wires, i.e., we consider the Coulomb
  drag geometry.  We show that the fluctuations in $G_{12}$ differ
  dramatically from those of the diagonal conductance $G_{ii}$: the
  fluctuations are large, and can even exceed the mean value, thus
  implying a possible reversal of the induced drag current.  We report
  extensive numerical simulations elucidating the fluctuations, both
  for correlated and uncorrelated disorder.  We also present analytic
  arguments, which fully account for the trends observed numerically.
\end{abstract}

\narrowtext

\section{Introduction}\label{introduction}

Recent advances in nanotechnology have made transport studies of
quantum wires an active field of study. Several fabrication routes
have become available: gated two-dimensional electron gases, cleaved
layer overgrowth techniques, grooved high-index surfaces, or even
nanotubes.  It is quite feasible that in future's nanoelectronic
components several quantum wires are very closely spaced (spacing $d$
is of the order of the inverse screening wave-vector), and hence the
stray fields due to moving charges in one quantum wire affect the
motion of charges in the neighboring wires. Analogous effects in
coupled quantum wells have been intensively studied for the last ten
years (for a review, see Ref. \onlinecite{rojo99}), and have become
known as Coulomb drag. While only very few experimental studies of
coupled quantum wires have been reported so far (however, see Refs.
\onlinecite{debray,yama2001}), we believe that more experiments will
become available in the near future.

Coulomb drag of mesoscopic structures has been addressed theoretically
in the case of 1D systems both within the Boltzmann equation
approach\cite{hu96,gure98} and for Luttinger liquids with strong
interwire
interactions.\cite{naza98,flen98,pono00,komn98,komn00,kles99} For the
latter case, the interesting possibility of a regime with almost
identical currents in the two wires has been
predicted\cite{naza98,flen98,pono00,komn98} and also interesting
effects for drag between carbon nanotubes have been
found.\cite{komn98,komn00}

Disorder will inevitably be present in all real samples. The study of
fluctuation phenomena in mesoscopic systems is a mature field, and has
tremendously increased our understanding of the basic physics
governing electronic transport in solids.  The observation and
explanation of {\it universal conductance fluctuations}\cite{feng88}
is one of the central achievements in this field. It is then natural
to ask: are similar phenomena present in the case of Coulomb coupled
systems? In other words, what are the fluctuation properties of the
{\it transconductance}?

\begin{figure}
\begin{center}
\epsfig{file=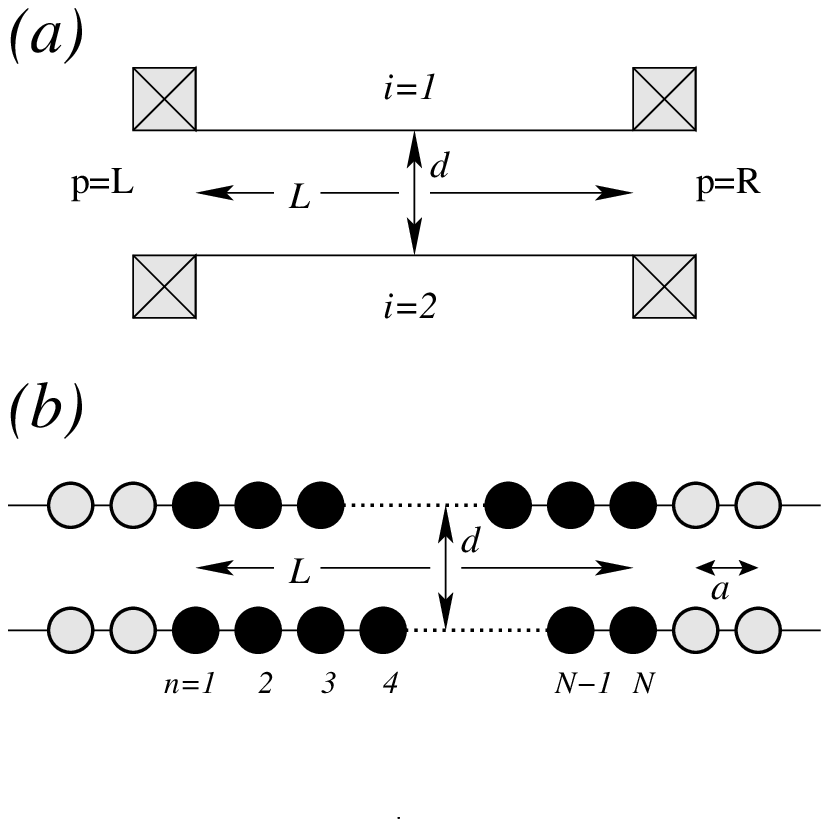, width=0.8\columnwidth,clip}
\end{center}
\caption{(a) Geometry of the two coupled 1D--wires of length $L$
and separation $d$. (b) Lattice model of the two 1D--wires. Here,
$\bullet$ denote the lattice points of the wires (where Coulomb
interaction $U_{12}$ is effective) and $\circ$ denote those
belonging to the ideal leads.}
\label{fig:sample}
\end{figure}

Quite recently the study of fluctuations of the Coulomb drag was
initiated by Narozhny and Aleiner.\cite{naro00} These ideas were used
by us to study various properties of Coulomb drag in systems smaller
than the Thouless energy.\cite{mort2001a,mort2001b} They provided
answers to some aspects to the question phrased above.  In particular,
in Ref.~\onlinecite{naro00} it was shown that the fluctuations will be pronounced for temperatures
smaller than the Thouless energy. In contrast to the universal
conductance fluctuations\cite{feng88} of the Landauer conductance
$G_{ii}$ the fluctuations of $G_{21}$ are however non-universal and
depend on temperature and various system parameters.

In this work we focus on coupled quasi-ballistic 1D wires (see
Fig.~\ref{fig:sample}) and show that even weak disorder can change the
ballistic properties drastically, and give rise to new interesting
phenomena. The paper also gives a more detailed derivation of some
technical results used in our recent Letter.\cite{mort2001a}

We consider temperatures smaller than $\hbar v_F/L$ so that
  intra-wire interactions are effectively given by the Fermi liquid
  reservoirs and where the screening can be considered to be static.
  In this regime it is well known from calculation of drag between
  clean Luttinger liquids that the intra-wire interactions have no
  pronounced effect.\cite{naza98,flen98} For the inter-wire
  interaction we consider the physical relevant regime with
  longe-range interaction where $U_{12}(q=0)\gg U_{12}(q=2k_F)$.

Starting from the Kubo formula we derive an expression for the
transconductance $G_{21}$ similar to previous
results\cite{kame95,flen95b} but relaxing the assumption of
translational invariance.\cite{mort2001a} We formulate the problem
with the help of the spectral function and subsequently map it onto a
tight-binding-like model suitable for a computer implementation. Our
formalism allows us to treat wires with electrons propagating in
arbitrary potentials. The fluctuations in drag between disordered
wires is studied by numerical ensemble averaging and also analytically
with the disorder included perturbatively. We consider two situations:
uncorrelated and correlated disorder. For the situation where the
disorder potentials of the two wires are mutually uncorrelated, as is
usually assumed,\cite{naro00,mort2001a,flen95b} we find large
fluctuations of the order of the mean value and thereby also a
possible sign reversal of the drag current. If the two disorder
potentials are identical, {\it i.e.} mutually correlated,\cite{gorn99}
we get large fluctuations, and an enhanced mean value compared to
uncorrelated disorder. We also give some results for the intermediate
case. The case of fully correlated disorder is in qualitative
agreement with recent predictions for 2D systems\cite{gorn99} and for
the fluctuations we predict an enhancement by a factor of $\sqrt{2}$
which is confirmed numerically.

The obtained distributions are quite robust in the sense that systems
with different disorder strength and/or length and separation can be
rescaled to fall on the same curve. However, the distributions also
depend on the range of the interaction $U_{12}$ and in that sense they
are non-universal.

The paper is organized as follows: in Sec.~\ref{formalism} we
introduce the formalism and in Sec.~\ref{analytics} we derive analytic
results for quasi-ballistic wires. Sec.~\ref{numerics} contains our
numerical results. Finally, in Sec.~\ref{conclusion} discussions and
conclusions are given. Certain technical details can be found in
appendices.

\begin{figure}
\begin{center}
\epsfig{file=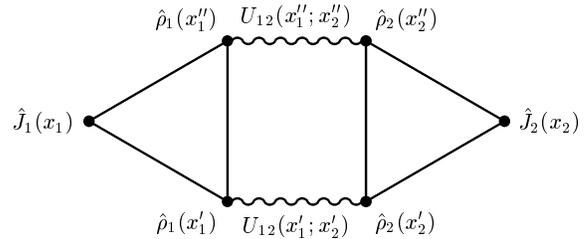, width=0.9\columnwidth,clip}
\end{center}
\caption{Diagrammatic representation of the current-cur\-rent correlation function
$\Delta_1 U_{12} U_{12}\Delta_2$, see Eq.~(\ref{G21}).}
\label{fig:triangle}
\end{figure}

\section{Formalism}\label{formalism}

\subsection{General formulation}

Let us consider two 1D--wires of length $L$ (shorter than the
phase-coherence length $\ell_\phi$) parallel to each other with a
separation $d$, see Fig.~\ref{fig:sample}. Our aim is to calculate the
dc ($\Omega\rightarrow 0$) linear-response transconductance
$G_{21}=\partial I_2/\partial V_1$. In the 1D situation the
transconductance is equal to the transconductivity which can be
calculated from Kubo formalism\cite{mahan} which expresses $G_{21}$ in
terms of the retarded current-current correlation
function\cite{flen95b}
\begin{eqnarray}
G_{21}&=&\lim_{\Omega\rightarrow 0}\frac{e^2}{\hbar\Omega}
 \int_{0}^\infty{\rm d}(t-t') \nonumber\\
&&\quad\times \exp\big[i\Omega(t-t')\big]
\big<\big[\hat{J}_2(x_2,t),\hat{J}_1(x_1,t')\big]\big>.
\end{eqnarray}
Here, $\hat{J}_i$ is the particle current operator of wire $i=1,2$.
The brackets indicate the quantum mechanical statistical average. Due
to current conservation $x_1$ and $x_2$ can be chosen at any position
along the wires.\cite{bara89,nock93} Using Matsubara formalism we
follow Ref.~\onlinecite{flen95b} and calculate $G_{21}$ to second
order in the interaction $U_{12}$ between the mesoscopic 1D--wires.
Intrawire interactions are neglected, except for disorder potential
scattering. The expansion of the time-development operator to second
order in $U_{12}$ gives rise to a current-current correlation function
which can be expressed as a product of two three-point correlation
functions: $\big<\hat\rho_1\hat\rho_1\hat{J}_1
\big>U_{12}U_{12}\big<\hat\rho_2\hat\rho_2\hat{J}_2 \big>$, where
$\hat\rho_i$ is the particle density operator. The three-point
correlation functions are evaluated with the aid of Wick's theorem and
after some lengthy, but in principle straight forward calculations, we
arrive at
%\wideequationbegin
\widetext
\begin{eqnarray}\label{G21}
&&G_{21}= \frac{e^2}{h}\int_0^L\int_0^L\int_0^L\int_0^L
{\rm d}x_1' {\rm d}x_2' {\rm d}x_1'' {\rm d}x_2''
\int_{-\infty}^\infty \hbar {\rm d}\omega\,
\frac{\Delta_1(\omega,x_1',x_1'') U_{12}(x_1';x_2')U_{12}(x_1'';x_2'')
\Delta_2(\omega,x_2'',x_2')}{2 kT\sinh^2(\hbar\omega/2kT)},\\
\label{JJcorrelator}
&&{\Delta}_i(\omega;x',x'')= \frac{1}{4\pi} \frac{\hbar^2}{2m}
\int{\rm d}\varepsilon\,
A_{\varepsilon-\hbar\omega}^i(x'',x')
\big[ A_{\varepsilon}^i(x,x'')\partial_{x} A_{\varepsilon}^i(x',x)
-A_\varepsilon^i(x',x)\partial_{x}A_{\varepsilon}^i(x,x'') \big]\nonumber\\
&&\quad\quad\quad\quad\quad\quad\quad\quad
\quad\quad\quad\quad\quad\quad\quad\quad\quad\quad\quad\times
\big[n_F(\varepsilon-\hbar\omega)-n_F(\varepsilon)\big]
-\big(\;\omega\rightarrow-\omega\;\big)^*,\\\label{A}
&&A_\varepsilon^i(x,x')=i\left[{\cal G}_\varepsilon^i(x,x')
-\big\{{\cal G}_\varepsilon^i(x',x)\big\}^*\right]
=2\pi \sum_\alpha \big<x\big|\alpha\big>\big<\alpha\big|x'\big>
\delta(\varepsilon-\varepsilon_\alpha).
\end{eqnarray}
%\narrowtext
\wideequationend
\noindent Here, $A_\varepsilon^i$ is the spectral function, ${\cal
  G}_\varepsilon^i$ the retarded Green function, $n_F(\varepsilon)$
the Fermi--Dirac distribution function, and $\varepsilon_\alpha$ is
the eigenvalue of the exact single-particle eigenstate
$\big|\alpha\big>$ in the uncoupled wires. We furthermore
assume spin-degeneracy. The current-current correlation function
$\Delta_1U_{12}U_{12}\Delta_2$ is shown schematically in
Fig.~\ref{fig:triangle}.

Eq.~(\ref{G21}) generalizes the results of Ref.~\onlinecite{flen95b}
to broken translational invariance, and App.~\ref{Delta:app}
establishes a connection to the expression employed in our recent
letter.\cite{mort2001a}

\subsection{Lattice formulation}

Writing the four spatial integrals in Eq.~(\ref{G21}) as sums over the
functions on a discrete lattice, see Fig.~\ref{fig:sample}, we get

\begin{eqnarray}\label{G21lattice_general}
G_{21}&=& \frac{e^2}{h}\int_{-\infty}^\infty\hbar {\rm d}\omega\,
\frac{{\rm Tr}
\big[ U_{12}^T \Delta_1(\omega) U_{12}\Delta_2(\omega) \big]}
{2 kT\sinh^2(\hbar\omega/2kT)},
\end{eqnarray}
where the matrices have elements
\begin{eqnarray}
&&\big\{U_{12}\big\}_{nn'}=U_{12}(x_1'\rightarrow na;x_2'\rightarrow n'a),\\
&&\big\{\Delta_{i}(\omega)\big\}_{nn'}
= \frac{1}{4\pi} \frac{\hbar^2}{2ma^2}\int{\rm d}\varepsilon
\big\{A_{\varepsilon-\hbar\omega}^i\big\}_{n'n} \nonumber\\
&&\quad\times
\big[ \big\{A_{\varepsilon}^i\big\}_{\tilde{n} n'}
\big\{A_{\varepsilon}^i\big\}_{n,\tilde{n}+1}
- \big\{A_{\varepsilon}^i\big\}_{n\tilde{n}}
\big\{A_{\varepsilon}^i\big\}_{\tilde{n}+1,n'} \big]
\nonumber\\
&&\quad\times
\big[n_F(\varepsilon-\hbar\omega)
-n_F(\varepsilon)\big]-\big(\;\omega\rightarrow-\omega\;\big)^*,\\
&&\big\{A_\varepsilon^i\big\}_{nn'}
\equiv a\times A_\varepsilon^i(x\rightarrow na,x'\rightarrow n'a),
\end{eqnarray}
where $n,n'=1,2,3,\ldots N$ label the lattice points and $a$ is the
lattice constant, see Fig.~\ref{fig:sample}(b). In the matrix notation
$A_\varepsilon^i=i\big[ {\cal G}_\varepsilon^i-\big\{{\cal
  G}_\varepsilon^i\big\}^\dagger\big]$. The derivative $\partial_{x}$
in $\Delta$ has been accounted for by the method of finite
differences\cite{datta} and $\tilde{n}$ can be any lattice points
$1,2,3,\ldots N-2,N-1$ due to current conservation.  Summing over the
first $N-1$ lattice points and dividing by $N-1$ we find
\begin{eqnarray}
&&\Delta_{i}(\omega)= \frac{1}{4\pi} \frac{\hbar^2}{2ma^2}
\int{\rm d}\varepsilon \big\{A_{\varepsilon-\hbar\omega}^i\big\}^T
\otimes\big[ A_{\varepsilon}^i\Lambda  A_{\varepsilon}^i \big]
\nonumber\\
&&\quad\times
\big[n_F(\varepsilon-\hbar\omega)-n_F(\varepsilon)\big]
-\big(\;\omega\rightarrow-\omega\;\big)^*,
\end{eqnarray}
where $\otimes$ denotes an element-by-element multiplication, $\{X
\otimes Y\}_{nm} = X_{nm}Y_{nm}$ and the matrix $\Lambda$ has elements
\begin{eqnarray}\label{Lambda}
\Lambda_{nn'}= \frac{\pm\delta_{n,n'\pm 1}}{N-1}.
\end{eqnarray}

The next step is to calculate the lattice representation of the
retarded Green functions ${\cal G}^1$ and ${\cal G}^2$ of the two
uncoupled wires. Writing the Laplacian $\partial_x^2$ in the wire
Hamiltonians with the help of finite differences the problem of the
uncoupled wires is mapped onto tight-binding-like Hamiltonians\cite{datta}

\begin{equation}
H_{nn'}^{(i)}=(2t+\{V_i\}_n)\delta_{nn'} -t\delta_{n,n'\pm 1}\,,\, t=\hbar^2/2ma^2.
\end{equation}
A standard approach based on Dyson's equation then gives the
retarded Green functions of the uncoupled wires as an $N\times N$ matrix\cite{datta}
\begin{equation}\label{green}
{\cal G}_\varepsilon^i=\big[\varepsilon - H^i
-\Sigma_{\rm \scriptscriptstyle L}^i(\varepsilon)
-\Sigma_{\rm \scriptscriptstyle R}^i(\varepsilon)\big]^{-1},
\end{equation}
where the couplings to lead $p={\rm L,R}$ are taken care of by the
retarded self-energy
\begin{equation}
\big\{\Sigma_{\scriptscriptstyle p}^i(\varepsilon)\big\}_{nn'}=
-t \exp(i k(\varepsilon) a)\,\delta_{n,n_p}\delta_{n_p,n'}
\end{equation}
with $n_{\rm L}=1$ and $n_{\rm R}=N$. The wave vector is related to
the energy through the usual cosine dispersion relation with a
band-width of $4t$, {\it i.e.} $\varepsilon=2t(1-\cos ka)$.

\subsection{Low temperature expansion}

We now consider the low temperature limit $kT\ll\varepsilon_F$ where
we can evaluate the spectral functions in Eq.~(\ref{JJcorrelator}) at
the Fermi level. Performing the energy integration gives
$\Delta_i\propto\omega$ and the $\omega$ integration in
Eq.~(\ref{G21}) can now be done:
\begin{equation}\label{T2}
\int_{-\infty}^\infty \hbar {\rm d}\omega\,
\frac{(\hbar\omega)^2}{2 kT\sinh^2(\hbar\omega/2kT)}
= \frac{4\pi^2}{3}\left(kT\right)^2,
\end{equation}
and we get $G_{21}\propto T^2$. Eq.~(\ref{G21lattice_general}) now
simplifies to

\begin{equation}
G_{21}=\frac{e^2}{h}\left(kT\right)^2 \frac{t^2}{3}
{\rm Tr}\big[U_{12} M_1\,U_{12}\,M_2\big],\label{G21lattice}
\end{equation}
where

\begin{eqnarray}\label{M}
M_i&=&{\rm Re}\big\{ \big\{A_{\varepsilon_F}^i\big\}^T\otimes
\big[ A_{\varepsilon_F}^i\Lambda A_{\varepsilon_F}^i \big]\big\}.
\end{eqnarray}
Eq.~(\ref{G21lattice}) forms the
basis for all subsequent numerical work.

The Landauer conductance $G_{ii}=\partial I_i/\partial V_i$ of the
individual wires can be expressed in a similar form\cite{datta}

\begin{equation}\label{Giilattice}
G_{ii}=\frac{2e^2}{h}{\rm Tr}
\big[\Gamma_{\rm \scriptscriptstyle L}^i(\varepsilon_F)
{\cal G}_{\varepsilon_F}^i
\Gamma_{\rm \scriptscriptstyle R}^i (\varepsilon_F)
\big\{{\cal G}_{\varepsilon_F}^i\big\}^\dagger \big],
\end{equation}
where the leads are described by
\begin{equation}
\Gamma_{\scriptscriptstyle p}^i(\varepsilon)
=i\big[\Sigma_{\scriptscriptstyle p}^i(\varepsilon) -
  \big\{\Sigma_{\scriptscriptstyle p}^i(\varepsilon)\big\}^\dagger\big].
\end{equation}

The matrix formulation is readily implemented on a computer and the
accuracy can be increased simply by having more lattice points $N$
(for a given $L$).\cite{datta} The retarded Green function can be
obtained either by a direct inversion as indicated in
Eq.~(\ref{green}) or by a recursive method, see {\it e.g.}
Ref.~\onlinecite{bara91}.

Alternatively one can also view Eqs.~(\ref{G21lattice}) and
(\ref{Giilattice}) as formulas for a tight-binding system where $t$ is
a 'hopping matrix element' between different orbitals and the local
potential $V_n+2t$ is the energy of the orbital localized at site $n$.
Finally we note that the form of Eq.~(\ref{G21lattice}) is valid for
any number of transverse channels (for 2D and 3D systems $\Lambda$ in
Eq.~(\ref{Lambda}) should be modified to also contain an 'integration'
over the transverse direction in order to get the total current) and
thus provides a versatile starting point for investigations of drag in
different geometries.

\subsection{Anderson model}

For the study of disordered wires we use the Anderson model with
diagonal disorder\cite{ande58} where the site potentials $V_n^i$ of a
given wire are statistically independent with each site energy taken
from a uniform distribution of width $W$ and zero mean (see also Ref.
\onlinecite{kram93}). The back-scattering mean free path $\ell$ can be
related to the disorder strength $W$ and in the Born approximation we
get (see App.~\ref{ell})

\begin{equation}\label{ell_F}
\ell=12a(4t\varepsilon_F - \varepsilon_F^2)/W^2.
\end{equation}

We will examine several different ways of realizing the disorder in
the two wires. These include a situation where both wires are
disordered, but with $V_1$ and $V_2$ mutually uncorrelated (as in
Ref.~\onlinecite{mort2001a}). We also present result for disordered
wires where the disorder is fully correlated, {\it i.e.} $V_1=V_2$ (as
suggested in Ref.~\onlinecite{gorn99}).

\section{Analytical results}\label{analytics}

In this section we consider quasi-ballistic wires. For weak disorder
it is possible to make a perturbation expansion for the
fluctuations\cite{naro00} $\delta G_{21}=G_{21}- \left<G_{21}\right>$
(brackets indicate ensemble averaging).

\begin{figure}
\begin{center}
\epsfig{file=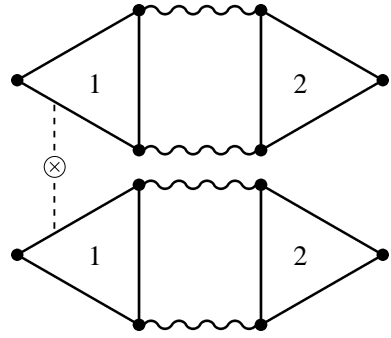, width=0.59\columnwidth,clip}
\end{center}
\caption{Example of a first order connected diagrams in the diagrammatic
expansion for the fluctuations giving rise to a
$\big<[\delta G_{21}]^2\big>\propto V^2\propto 1/k_F\ell$ dependence.
Due to momentum conservation the impurity line must carry $q=0$ and the four interactionlines $2k_F$. }
\label{fig:fluctuations_first}
\end{figure}

\subsection{Fluctuations of the order of $V^2$}
An example of a lowest order connected fluctuation diagram is shown in
Fig. \ref{fig:fluctuations_first}. Due to momentum conservation the
impurity line must carry $q=0$ (corresponding to forward-scattering)
and the four interactionlines $2k_F$ so that we estimate the magnitude
as follows

\begin{equation}\label{forward}
\big<[\delta G_{21}]^2\big> \propto V^2(0) U_{12}^4(2k_F),
\end{equation}
where $V(q)$ and $U_{12}(q)$ are the Fourier transforms of the
disorder potential and the Coulomb interaction potential,
respectively. Though diagrams of order $V^4$ are parametrically smaller, we shall see that they give the dominant contribution in the case of
long-range Coulomb interaction.

\subsection{Fluctuations of the order of $V^4$}
\begin{figure}
\begin{center}
\epsfig{file=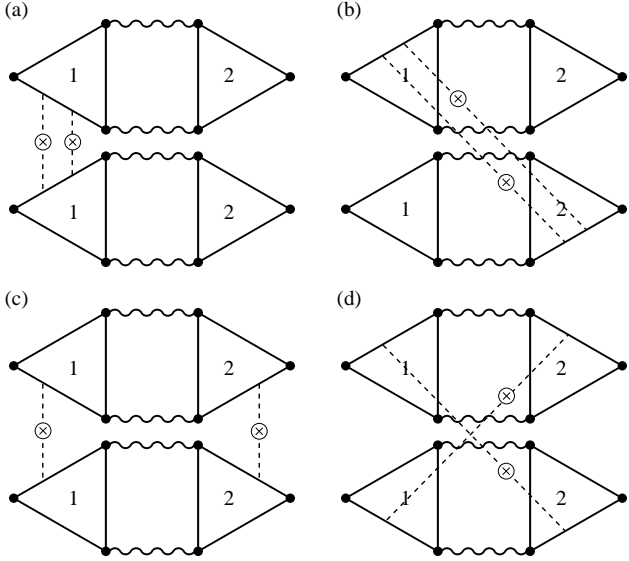, width=0.99\columnwidth,clip}
\end{center}
\caption{Examples of second order connected diagrams in the diagrammatic
expansion for the fluctuations giving rise to a
$\big<[\delta G_{21}]^2\big>\propto V^4\propto 1/(k_F\ell)^2$ dependence.
Due to momentum conservation diagrams with only one impurity line do not
contribute to the drag fluctuations. The diagrams (a) and (c)
are relevant for both correlated and uncorrelated disorder whereas the diagrams
in panels (b) and (d) are relevant for correlated disorder only. }
\label{fig:fluctuations_diagrams}
\end{figure}

Fig.~\ref{fig:fluctuations_diagrams} shows examples of connected
diagrams of the order of $V^4$; now the impurity lines can carry both
$q=0$ and $q=2k_F$ corresponding to back-scattering. For
back-scattering in the disorder channel we in the case of
un-correlated ($\rm uc$) disorder get the estimates

\begin{subequations}\label{V4-contributions}
\begin{eqnarray}
a:&& \big<[\delta G_{21}]^2\big>_{\rm uc} \propto V^4(2k_F) U_{12}^4(2k_F) ,\\
b:&& \big<[\delta G_{21}]^2\big>_{\rm uc} =0,\\
c:&& \big<[\delta G_{21}]^2\big>_{\rm uc} \propto V^4(2k_F) U_{12}^2(0) U_{12}^2(2k_F),\label{fluc_c}\\
d:&& \big<[\delta G_{21}]^2\big>_{\rm uc} =0,
\end{eqnarray}
\end{subequations}
where to lowest order $V^2(2k_F)\propto 1/k_F \ell$ so that 
\begin{equation}\label{1/ell}
\big<[\delta G_{21}]^2\big>^{1/2} \propto  1/k_F \ell.
\end{equation}

For a long-range Coulomb potential $U_{12}(0)\gg U_{12}(2k_F)$ the
dominant contribution comes from diagram $c$ which also dominates
parametrically over the contribution to $2$nd order, see
Eq.~(\ref{forward}). Screening mainly affects the small-$q$ limit of
$U_{12}(q)$. The mean value $\big<G_{21}\big>\propto U_{12}^2(q=2k_F)$
is thus only weakly affected by screening whereas the
$V^4$-contributions to the fluctuations will be strongly suppressed
due to the presence of also $U_{12}^2(q=0)$, see
Eqs.~(\ref{V4-contributions}). For sufficiently short-range
interaction the $V^2$-contributions will eventually give the dominant
contribution, see Eq.~(\ref{forward}), but then the overall magnitude
of the fluctuations will also be small compared to the mean value.

For correlated disorder ($\rm c$) all diagrams $a$, $b$, $c$ and $d$
contribute (equally) whereas for uncorrelated disorder only the
diagrams $a$ and $c$ are relevant. More generally, for each
topologically different diagram contributing in the case of
uncorrelated disorder there are two similar diagrams contributing
equally in case of correlated disorder. Of course there are also other
possible diagrams in case of correlated disorder but to lowest order
in $1/k_F\ell$ this means that

\begin{eqnarray}\label{sqrt2}
\frac{\big<[\delta G_{21}]^2\big>_{\rm c}^{1/2}}{\big<[\delta G_{21}]^2\big>_{\rm uc}^{1/2}}
&\simeq& \sqrt{\frac{\big<a\big>+\big<b\big>+\big<c\big>+\big<d\big>}{\big<a\big>+\big<c\big>}}\nonumber\\
&=& \sqrt{\frac{2\times \big<a\big>+2\times\big<c\big>}{\big<a\big>+\big<c\big>}}=\sqrt{2},
\end{eqnarray}
where $\big<a\big>$, $\big<b\big>$, $\big<c\big>$, and $\big<d\big>$
refer symbolically to the diagrams in
Fig.~\ref{fig:fluctuations_diagrams} averaged over disorder. This
simple argument suggests that the fluctuations in the case of
correlated disorder will be enhanced by a factor of $\sqrt{2}$
compared to the case of uncorrelated disorder.

We now turn into a quantitative evaluation of the fluctuations. In
App.~\ref{alternative} we show how $\Delta_i(x,x')$ can be expressed
in terms of scattering states $\psi_\pm (x)$. We introduce a
short-hand notation

\begin{equation}
G_{21}={\cal C}\int  U_{12}(x,y)U_{12}(x'y') \Delta_1(x,x')\Delta_2(y',y),
\end{equation}
with
\begin{equation}
\Delta(x,x')= {\rm Im}\big\{\tilde\rho_{+-}(x)\tilde\rho_{-+}(x')\big\},
\end{equation}
where $\int$ means an integral over all spatial degrees of freedom (in
this case $x$, $x'$, $y$, and $y'$). Here, $\tilde\rho$ are the particle density matrix elements taken in a new basis $\tilde\psi_\pm
(x)$ related to the original basis $\psi_\pm
(x)$ by a unitary
transformation described in App.~\ref{alternative}. The frequency
integration has been performed to give the $T^2$-dependence and all
prefactors are included in ${\cal C}$, {\it i.e.} ${\cal C}\propto
(e^2/h)(kT)^2$.

Before calculating the prefactor in Eq.~(\ref{1/ell}) in the case of
uncorrelated disorder we first consider the transconductance in the
case of ideal ballistic wires. Using the free plane-waves $ \tilde\psi_\pm
(x)=\psi_\pm (x)=e^{\pm i k_fx}$ we get

\begin{equation}
\bar\Delta(x,x')= {\rm Im}\big\{e^{-i2k_Fx}e^{i2k_Fx'} \big\}=\sin2k_F(x'-x).
\end{equation}
The electrons in each wire can be either forward-scattered (zero
momentum) or back-scattered ($2k_F$) and because of momentum
conservation both the interaction lines in the diagram in
Fig.~\ref{fig:triangle} carry momentum $2k_F$ (zero momentum from
forward scattering does not contribute to drag in this situation) so
that

\begin{equation}\label{2kF}
G_{21}(\infty) = {\cal C} \frac{1}{2} U_{12}^2(2k_F),
\end{equation}
where
\begin{equation}
U_{12}(q)=\int_0^L\int_0^Ldx_1dx_2\,e^{iq(x_1-x_2)}U_{12}(x_1,x_2),
\end{equation}
In what follows we often express the results for disordered systems
normalized with respect to the ballistic transconductance
$G_{12}(\infty)$.

The situation is very different even for weak disorder, and the broken
translational invariance allows for transferred momentum others than
$2k_F$. For the diagrams in Fig.~\ref{fig:fluctuations_diagrams} the
impurity lines carry momentum $2k_F$ corresponding to back-scattering
within the wires (forward-scattering will not contribute to the
fluctuations).  However, the difference between diagram (a) and (c) is
that in (a) the interaction lines in each sub-diagram must carry the
same momentum whereas in (c) one of them can carry {\it e.g.} zero
momentum while the other carries $2k_F$. Now, since $U_{12}(0)\gg
U_{12}(2k_F)$ for long-ranged interaction (typically, $U_{12}(q)$
decays exponentially) the diagram (c) gives the major
contribution to the fluctuations. This can be tested numerically by
noting that the diagram (a) is relevant to the case where
wire $1$ is disordered and wire $2$ is either disordered or ballistic
whereas the diagram (c) is only relevant to the case of both
wires being disordered. Indeed, by numerically calculating the
fluctuations for a system where one of the wires is ballistic and the
other disordered we have found a very dramatic reduction (by more than
an order of magnitude) of the fluctuations compared to the case of
both wires being disordered.

Considering wires with mutually uncorrelated disorder and writing
$\Delta(x,x')=\bar\Delta(x,x')+\delta\Delta(x,x')$ (where the second
term is the correction to the ballistic limit) the contribution from
diagram (c) is

\begin{eqnarray}\label{deltaG2}
\big<[\delta G_{21}]^2\big>_{\rm uc}& \simeq&
{\cal C}^2 \int  U_{12}(x,y)U_{12}(x',y') U_{12}(\bar{x},\bar{y})\nonumber\\
&&\quad\times U_{12}(\bar{x}',\bar{y}')  \big<\delta\Delta_1(x,x')\delta
\Delta_1(\bar{x},\bar{x}')\big>\nonumber\\
&&\quad\big<\delta\Delta_2(y',y)\delta\Delta_2(\bar{y}',\bar{y})\big>,
\end{eqnarray}
and the remaining problem is to calculate the correlation
$\big<\delta\Delta_i(x,x')\delta\Delta_i(\bar{x},\bar{x}')\big>$ to
lowest order (second order) in the disorder strength.  For
quasi-ballistic wires (${\cal T}=1-{\cal R}\sim 1$)
Eqs.~(\ref{Delta_tilde}) and (\ref{rho_tilde}) give

\begin{equation}
\Delta(x,x')\approx {\rm Im}\big\{\rho_{+-}(x)\rho_{-+}(x')\big\},
\end{equation}
with $\rho_{\alpha\beta}(x)=\psi_\alpha^*(x)\psi_\beta(x)$ in the
original basis and formally we thus have that
\begin{equation}
\delta\Delta(x,x')= {\rm Im}\big\{\delta \rho_{+-}(x)e^{i2k_Fx'}+
e^{-i2k_Fx}\delta \rho_{+-}^*(x') \big\}.
\end{equation}
Using the Lippmann--Schwinger equation\cite{lippmann-schwinger} we have to
lowest order in the disorder strength that

\begin{equation}
\psi_\pm(x)\simeq e^{\pm i k_Fx}+\int_0^L d\chi\,{\cal G}_0(x,\chi)V(\chi)
e^{\pm i k_F\chi},
\end{equation}
where ${\cal G}_0(x,x')=(i\hbar v_F)^{-1}e^{ik_F|x'-x|}$ is the
unperturbed retarded Green function. This means that $\rho_{+-}(x)= \rho_{+-}^0(x) + u(x)+\nu(x)$ where

\begin{eqnarray}
u(x)
&=& (i\hbar v_F)^{-1}\int_0^L d\chi\,{\rm sign}(x-\chi)V(\chi)\rho_{+-}^0(\chi),\\
\nu(x)&=&(i\hbar v_F)^{-1}\int_0^L d\chi\,
{\rm sign}(\chi-x)V(\chi)\rho_{+-}^0(x).
\end{eqnarray}
Here, $\rho_{+-}^0(x)=\exp(-2ik_Fx)$. Thus we see that $u$ gives a non-oscillating correction to  $\rho_{+-}^0(x)$ which is the reason why the Fourier component $U_{12}(0)$ enters the end-result instead of $U_{12}(2k_F)$. This is not the case for $v$ and we can therefore omit it. For the averages over disorder we use
$\big<V(\chi)V(\chi')\big>=W_0^2\delta(\chi-\chi')$ and neglecting
terms oscillating with $4k_F$ we get $\big<u(x)u(\bar{x})\big>\approx
0$ and

\begin{equation}
\big<u(x)u^*(\bar{x})\big>=L\left(\tfrac{U_0}{\hbar v_F}\right)^2
\big(1-2\big|x-\bar{x}\big|/L\big).\nonumber
\end{equation}
The prefactor can be related to the reflection coefficient of the
wire. To see this we consider the Dyson equation to second order in
the disorder

\begin{multline}
{\cal G}(x,x')\simeq {\cal G}_0(x,x') +\int_0^L d\chi\,{\cal G}_0(x,\chi)V(\chi)
{\cal G}_0(\chi,x')\\
+\int_0^L d\chi\int_0^L d\chi'\,{\cal G}_0(x,\chi)V(\chi){\cal G}_0(\chi,\chi')
V(\chi'){\cal G}_0(\chi',x'),
\end{multline}
and from the Fisher--Lee relation we notice that ${\cal R}=1-{\cal T}=
1- (\hbar v_F)^2\big|{\cal G}(L,0)\big|^2$. It then follows that to
second order in the disorder $\big<{\cal R}\big> \simeq
L\left(W_0/\hbar v_F\right)^2=L/\ell$ (the last equality follows from
Appendix ~\ref{ell}) so that

\begin{equation}
\big<u(x)u^*(\bar{x})\big>=\big<{\cal R}\big>\big(1-2\big|x-\bar{x}\big|/L\big).
\end{equation}
This means that
\begin{eqnarray}
\big< \delta\Delta(x,x')\delta\Delta(\bar{x},\bar{x}')\big>
&=&\frac{\big<{\cal R}\big>}{2}\big\{f(x'-\bar{x}',x-\bar{x})\nonumber \\
&&\quad +f(x-\bar{x},x'-\bar{x}')\nonumber\\
&&\quad -f(x'-\bar{x},x-\bar{x}')\nonumber\\
&&\quad - f(x-\bar{x}',x'-\bar{x}) \big\},
\end{eqnarray}
where
$f(x'-\bar{x}',x-\bar{x})=(1-2|x'-\bar{x}'|/L)\cos2k_F(x-\bar{x})$.

Out of the 16 only the 4 terms of the form
$\cos(x-\bar{x})\cos(y-\bar{y})$ are finite for long wires $k_FL\gg 1$
and we thus get

\begin{eqnarray}
\big<[\delta G_{21}]^2\big>_{\rm uc} &\simeq& {\cal C}^2  \big<{\cal R}_1\big>
\big<{\cal R}_2\big> \int  U_{12}(x,y)U_{12}(x',y')\nonumber\\
&&\quad\times  U_{12}(\bar{x},\bar{y})U_{12}(\bar{x}',\bar{y}')\nonumber\\
&&\quad\times f(x-\bar{x},x'-\bar{x}')f(y-\bar{y},y'-\bar{y}') \nonumber\\
&=& {\cal C}^2  \frac{1}{2}\big<{\cal R}_1\big> \big<{\cal R}_2\big> U_{12}^2(2k_F)
\widetilde{U}_{12}^2(0).
\end{eqnarray}
This corresponds to the estimate in Eq.~(\ref{fluc_c}). Finally, the
mean value is close the value for ballistic wires and normalizing by
Eq.~(\ref{2kF}) we get

\begin{eqnarray}\label{analytic}
\frac{\big<[\delta G_{21}]^2\big>_{\rm uc}^{1/2}}{G_{21}(\infty)}
&\simeq&  \frac{\big[2\big<{\cal R}_1\big> \big<{\cal R}_2\big> U_{12}^2(2k_F)
\widetilde{U}_{12}^2(0)\big]^{1/2}}{U_{12}^2(2k_F)}
\end{eqnarray}
where

\begin{eqnarray}&&\widetilde{U}_{12}^2(0)\equiv \int_{0}^{L}
    \int_{0}^{L}\int_{0}^{L}\int_{0}^{L} dx_{1}\,dx_{2}\,dx_{1}'\,dx_{2}'
   \nonumber\\
&&\quad\times U_{12}(x_{1},x_{2}) U_{12}(x_{1}',x_{2}')
\Big(1-\tfrac{2|x_{1}-x_{1}'|}{L}\Big)\Big(1-\tfrac{2|x_{2}-x_{2}'|}{L}\Big),\nonumber
\end{eqnarray}
and $\big<{\cal R}_i\big>=L/\ell$. Eq.~(\ref{analytic}) predicts the
$1/\ell$ dependence found in the qualitative discussion leading to
Eq.~(\ref{1/ell}). It follows that the relative magnitude of the
fluctuations is of the order $\big<{\cal
  R}\big>U_{12}(0)/U_{12}(2k_F)$ and even though $\big<{\cal
  R}\big>\ll 1$ for quasi-ballistic wires the effect of long-ranged
interaction $U_{12}(0)/U_{12}(2k_F)\gg 1$ can give rise to relative
fluctuations of order unity, {\it i.e.} fluctuations comparable to the
mean value. As we shall see in the next section, Eq.~(\ref{analytic})
is in excellent agreement with our numerical results.

\section{Numerical results}\label{numerics}

In this section we apply Eq.~(\ref{G21lattice}) to numerically
evaluate the statistical properties of ensembles of different disorder
configurations.

The numerical implementation was tested by first calculating the
Landauer conductance $G_{ii}$, Eq.~(\ref{Giilattice}), for various
geometries which can also be solved analytically. We have also tested
our use of the Anderson model for disordered wires. By changing the
ratio $\ell/L$ from the localized regime ($\ell/L\ll 1$) to the
de-localized regime ($\ell/L\gg 1$) we have obtained distributions,
mean values, and fluctuations for the Landauer conductance $G_{ii}$
which are in agreement with the analytical results of
Abrikosov.\cite{abri81} We have also compared our lattice
implementation, Eq.~(\ref{G21lattice}) to analytical results, see
App.~\ref{ballistic}, for drag between two ballistic wires. By making
the grid sufficiently fine Eq.~(\ref{G21lattice}) is capable of
reproducing the curves in Fig.~\ref{fig:ballistic} which are based on
Eq.~(\ref{eq:ballistic}) with the integrals evaluated numerically.
Another test is the predicted peaking of $G_{21}$ at the onset of
modes in either of the two wires\cite{mort2001a} and for systems with
resonance transmission (peaks in $G_{ii}$); we find this to be born
out by our lattice implementation.

For the study of drag between disordered wires we consider
quarter-filled bands ($\varepsilon_F=t$) and wires with $N=100$
lattice points so that $k_FL =(\pi/3)\times 100$. The separation is
$k_F d= 1$ and for simplicity we assume unscreened Coulomb
interaction,

\begin{equation}\label{U(r)}
\big\{U_{12}\big\}_{nn'}=\frac{e^2}{4\pi \epsilon_0\epsilon_r\sqrt{
(n-n')^2 a^2 +d^2}}.
\end{equation}
For a discussion on how to include an effective screened interaction
({\it e.g.} within the random-phase approximation) see {\it e.g.}
Ref.~\onlinecite{flen95b} and references therein.

\begin{figure}[htb!]
\begin{center}
\epsfig{file=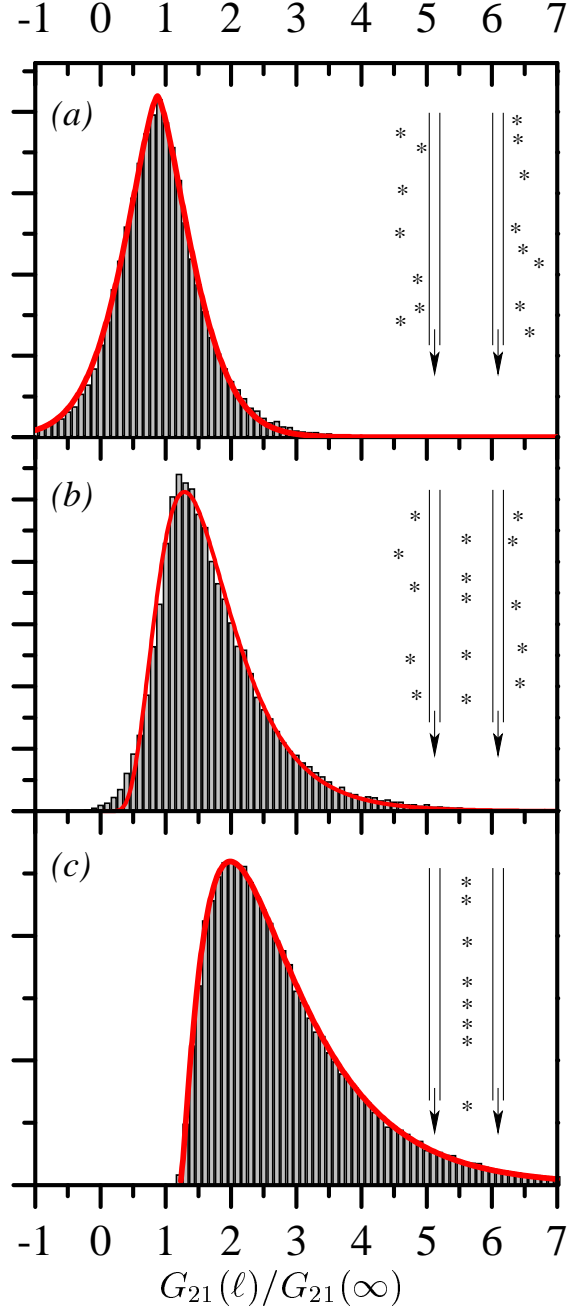, width=0.99\columnwidth,clip}
\end{center}
\caption{Histograms for $G_{21}(\ell)$ normalized by the ballistic result $G_{21}(\infty)$. The histograms are based on $>10^4$ disorder configurations and for all three histograms the mean free path is $\ell=36L$.
    Panel (a) is for the situation of mutually un-correlated disorder, panel (b) for partly correlated disorder ($W_{\rm c}=W_{\rm uc}$), and panel (c) is for mutually fully correlated disorder.}
\label{fig:histograms}
\end{figure}

For the case $\ell = 36L$ (this corresponds to $W=\varepsilon_F/10$)
the disorder has as expected\cite{abri81} almost no effect on the
Landauer conductance, {\it i.e.} $\big<G_{ii}\big>\sim 2e^2/h$ with
vanishing fluctuations. However, for the transconductance the
situation is very different. Panel (a) of Fig.~\ref{fig:histograms}
shows a typical histogram of $G_{21}(\ell)/G_{21}(\infty)$ (where
$G_{21}(\infty)$ is the result in the ballistic regime, see
App.~\ref{ballistic}) for $\ell =36 L$ in the case of uncorrelated
disorder. Depending on the disorder configuration $G_{21}(\ell)$ can
be either higher or lower

\noindent than in the ballistic regime. The
enhancement occurring for certain disorder configurations can be
understood physically as follows. The lack of translational invariance
allows forward scattering (transferred momentum $q\simeq0$), which
normally has little effect, to cause transitions between scattering
states with opposite directions, thus contributing to the drag.  We
emphasize that the histogram peaks close to the ballistic value and
not at zero drag, {\it i.e.} the mean drag is finite and positive. The
variance is of the same order as the mean value so that sign reversal
for some disorder realizations is possible. The later is represented
by the negative tail in the histogram. The sign of the drag is thus
arbitrary in the sense that both positive and negative drag can be
observed.

For the distribution of $G_{21}$ we find that $P_{\rm uc}(x)\propto
\exp[-|(x-\bar{x})/\tilde{x}|^\alpha]$ with $\alpha\simeq 1.4$ fits
surprisingly well to the data. However, this observation does not
exclude the possibility that other functions may be fitted equally
well.\cite{cheianov} In fact we have performed these fits to
histograms for $k_F\ell$ ranging from $10^3$ to $10^5$ with $k_FL$ in
the range $100$ to $300$ and find indications that $\alpha\sim
1.4\,-\, 1.5 $ which by rescaling of $G_{21}$ makes it possible to let
all histograms fall onto the same curve. However, the distribution is
non-universal in the sense that it depends on the range of the
interaction $U_{12}$.

\begin{figure}
\begin{center}
\epsfig{file=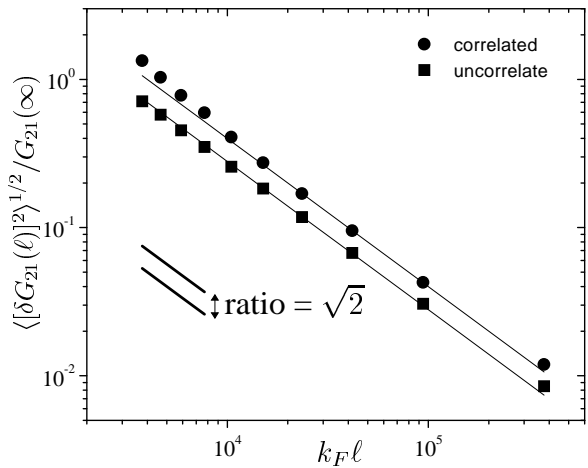, width=0.99\columnwidth,clip}
\end{center}
\caption{Plot of fluctuations $\big<[\delta G_{21}(\ell)]^2\big>^{1/2}$ as a
function of the mean free path $k_F\ell$. In both cases the expected $1/\ell$
behavior, Eq.~(\ref{1/ell}), is born out by the numerical calculations.
The full lines are Eq.~(\ref{analytic}) with no free parameters. The expected
enhancement, Eq.~(\ref{sqrt2}), for correlated disorder by a factor of $\sqrt{2}$
compared to uncorrelated disorder is also confirmed by the numerical calculations.}
\label{fig:fluctuations}
\end{figure}

For the same system parameters but now with fully correlated disorder
we get a very different distribution as seen in panel (c) of
Fig.~\ref{fig:histograms}. In this situation the mean transconductance
is enhanced compared to the uncorrelated case. This confirms at least
qualitatively the predictions for 2D systems of
Ref.~\onlinecite{gorn99}. The mean fluctuations are also enhanced (we
return to the enhancement below).  However, since the mean
transconductance is enhanced by almost a factor of two compared to the
ballistic limit this also means that there is no disorder
configurations giving rise to negative drag. For partly correlated
disorder, panel (b), the distribution is shifted to a higher mean
value and the shape is also slightly changed with an increase in the
magnitude of the fluctuations.

In Fig.~\ref{fig:fluctuations} we show the dependence of the
fluctuations on the mean free path $\ell$ which has the expected
$1/k_F\ell$ dependence, see Eq.~(\ref{1/ell}). We also notice that
correlated disorder give rise to slightly larger fluctuations compared
to uncorrelated disorder. In fact they are exactly enhanced by a
factor of $\sqrt{2}$ as predicted, see Eq.~(\ref{sqrt2}).

\section{Discussion and Conclusion}\label{conclusion}

We have studied drag of Coulomb coupled disordered mesoscopic
1D--wires and developed a formalism for investigating the statistical
properties both analytically and numerically. In this work we have
focused on the quasi-ballistic regime with $L \ll \ell$. For the
ordinary Landauer conductance $G_{ii}$ this is experimentally a
somewhat trivial regime since weak disorder has almost no effect and
$\big<G_{ii}\big>\sim 2e^2/h$ with vanishing fluctuations. However,
surprisingly this same regime (which can be achieved experimentally)
offers rich physics when measuring transconductance between two
Coulomb coupled wires. In this work we report results on the
statistical properties of the transconductance $G_{21}$ including its
distribution function.  Depending on the disorder configuration the
transconductance can be either positive or negative and even though
the mean value is close to the ballistic value the fluctuations can be
of the same order of magnitude. This is fully explained by analytical
calculations including the weak disorder perturbatively. The effect
can be explained by a combination of disorder induced back-scattering
in the wires and forward scattering between the wires induced by
Coulomb interaction.

We have also studied the recently proposed situation\cite{gorn99}
where the electrons in the two wires experience a common disorder
potential. In this case we also find that even weak disorder gives
rise to pronounced fluctuations and also a considerable enhancement of
the mean transconductance.

The reported distribution functions have been successfully fitted by
simple analytical expressions. Though there are no analytical
predictions for the exact functional form these initial results
suggest that a more detailed analytical investigation could be
rewarding.

Finally, let us note that while this work has focused on $L\ll \ell$,
for multi-mode wires also the diffusive and localized
regimes\cite{kram93,been97} with $L>\ell$ seems to be a promising
direction for future work.\cite{mort2001a}

\section*{Acknowledgement}

We thank C.~W.~J. Beenakker, M. Brandbyge, and V.~V. Cheianov for
useful discussions. N.A.M. acknowledges financial support by
``Ingeni\o rvidenskabelig Fond'', and ``G.A. Hagemanns Mindefond''.

\appendix

\section{Formulation in terms of spectral functions}\label{Delta:app}

We start from Eq.~(2) of Ref.~\onlinecite{mort2001a}. With a knowledge
of the exact eigenstates of the isolated wires the expression points
out a direct way of calculating the transconductance and it is
possible to simplify this expression for ballistic wires where the
eigenstates are known analytically, see App.~\ref{ballistic}. However,
in most cases the eigenstates are not known and it is useful to make a
numerically more appropriate formulation in terms of the retarded
lattice Green function. Using that the current matrix element can be
written as
\begin{equation}
\big<\alpha\big|\hat{J}(x)\big|\beta\big>=\frac{\hbar}{2mi}
\lim_{\tilde{x}\rightarrow x} \partial_{\tilde{x}}
\left[\big<\alpha\big|x\big>\big<\tilde{x}\big|\beta\big>
-\big<\alpha\big|\tilde{x}\big>\big<x\big|\beta\big> \right],
\end{equation}
and by introducing the factor $1=\int{\rm
  d}\varepsilon\,\delta(\varepsilon-\varepsilon_\alpha)$ into Eq.~(2)
of Ref.~\onlinecite{mort2001a} we get the result in
Eq.~(\ref{JJcorrelator}). In the second term we have used that the
interchange $x'\leftrightarrow x''$ corresponds to a sign-change along
with a complex conjugation.

\section{Mean Free Path in the Anderson Model}\label{ell}

For a tight-binding chain with $N$ sites and no disorder we have
eigenstates $\psi_k(n)=N^{-1/2} \exp(i k n a)$. We calculate the rate
for forward ($+$) and back ($-$) scattering from Fermi's golden rule

\begin{equation}
\frac{1}{\tau_\pm(k)}=\frac{2\pi}{\hbar}\sum_{k'} 
\left| \big< \psi_{k'} \big| V \big| \psi_k\big> \right|^2
\delta(\varepsilon-\varepsilon') \frac{1\pm kk'/|k||k'|}{2},
\end{equation}
where $V$ is the disorder potential which is treated as a
perturbation. For a chain with large $N$ we take the sum into an
integral and since $ \delta(\varepsilon-\varepsilon')=\left|\hbar
  v_k\right|^{-1}[\delta(k-k')+\delta(k+k')] $ we get
\begin{equation}
\frac{1}{\tau_\pm(k)}=\frac{a}{N\hbar^2 \left| v_k\right|}
\sum_{nm}e^{i(1\mp 1)k(n-m)a}\, V_n\,V_m .
\end{equation}

In the Anderson model\cite{ande58} the different sites are
uncorrelated and $p(V_n)=\Theta(W/2-V_n)/W$ which means that $\big<
V_n\,V_m\big> = \big< V_n^2\big>\delta_{nm}= W^2/12\, \delta_{nm}$.
The corresponding mean free path is given by $\ell(k)=v_k
/\big<{\tau_\pm^{-1}(k)}\big>$ (forward and back scattering give rise
to the same result) and for $\varepsilon=2t(1-\cos ka)$ we get the
result in Eq.~(\ref{ell_F}) at the Fermi level. The result agrees with
Ref.~\onlinecite{kram93} except for a constant shift of the energy by
$2t$.

\section{Ballistic regime}\label{ballistic}

For ballistic wires the eigenstates are of the form
$\psi_k(x)=L^{-1/2} \exp(ik x)$ with $\varepsilon=\hbar^2k^2/2m$. The
spectral function, Eq.~(\ref{A}), is then given by
\begin{equation}\label{A0}
A_\varepsilon(x,x')=k(\varepsilon) \cos [k(\varepsilon)(x'-x)]/\varepsilon.
\end{equation}
We consider the low temperature limit $kT\ll \varepsilon_F$ where we
can evaluate the spectral functions in Eq.~(\ref{JJcorrelator}) at the
Fermi level so that

\begin{eqnarray}\label{Delta_ballistic}
&&{\Delta}_i(\omega,x',x'')
= \frac{m}{2\pi\hbar^2}\frac{\hbar\omega}{\varepsilon_F}
\sin\left[2k_F\left(x'-x''\right)\right].
\end{eqnarray}
The $\omega$ integration in Eq.~(\ref{G21}) can now be performed and
using Eq.~(\ref{T2}) we get
\begin{eqnarray}\label{eq:ballistic}
&&G_{21}= \frac{e^2}{h}\frac{1}{12}\left(\frac{kT}{\varepsilon_F}\right)^2
\left(\frac{2m}{\hbar^2}\right)^2\nonumber\\&&
\quad\times \int_0^L\int_0^L\int_0^L\int_0^L {\rm d}x_1' {\rm d}x_2'
{\rm d}x_1'' {\rm d}x_2''\,
\sin\left[2k_F\left(x_1'-x_1''\right)\right]\nonumber\\&&
\quad\times  U_{12}(x_1';x_2')U_{12}(x_1'';x_2'')
\sin\left[2k_F\left(x_2'-x_2''\right)\right].
\end{eqnarray}

\begin{figure}
\begin{center}
\epsfig{file=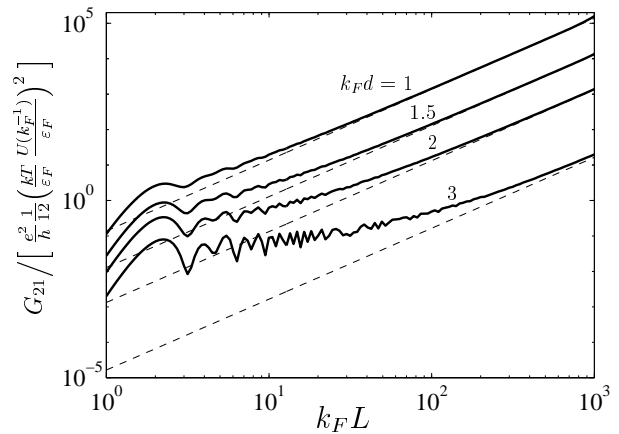, width=0.99\columnwidth,clip}
\end{center}
\caption{Plot of $G_{21}$ for two ballistic wires, Eq.~(\ref{eq:ballistic}), as a
function of the length $k_FL$ for different values of the separation $k_Fd$.
The dashed curves show the asymptotic result, Eq.~(\ref{eq:ballistic_assymp}).}
\label{fig:ballistic}
\end{figure}

In general the integrals have to be calculated numerically and only in
the limit $k_FL\gg 1$ (the limit also studied in
Ref.~\onlinecite{gure98}) we have the asymptotic result

\begin{equation}\label{eq:ballistic_assymp}
G_{21}\simeq \frac{e^2}{h} \frac{1}{6} \left(\frac{kT}{\varepsilon_F}\,
\frac{ U(k_F^{-1})}{\varepsilon_F}\, k_FL \, K_0(2k_Fd)\right)^2 ,
\end{equation}
where $K_0$ is a modified Bessel function of the second kind of order
zero.\cite{bessel} Here, we have assumed unscreened Coulomb
interaction, see Eq.~(\ref{U(r)}), and introduced
$U(r)=e^2/(4\pi\epsilon_0\epsilon_r r)$. Fig.~\ref{fig:ballistic}
shows a numerical evaluation of Eq.~(\ref{eq:ballistic}) along with
the asymptotic result, Eq.~(\ref{eq:ballistic_assymp}).

\section{Formulation in terms of scattering states}\label{alternative}

We start from Eq.~(4) in Ref.~\onlinecite{mort2001a} using scattering
states at the Fermi level as our basis: $\{\psi_+(x),\psi_-(x)\}$
where $+$ ($-$) is for a state incident from the left (right) lead.
All matrix elements $I_{\alpha\beta}$ and $\rho_{\alpha\beta}$ can
then be considered as elements of $2\times 2$ matrices: $I$ and
$\rho$. Due to current conservation we are free to evaluate the
current matrix element in the leads (outside the region with elastic
scattering) which gives $I=(v_F/ L)J$ with

\begin{eqnarray}
J&=&(\tau^3 -S^\dagger \tau^3 S)/2\nonumber\\&=& \left(
\begin{array}{cc}{\cal T} & -\sqrt{\cal RT}e^{i(\phi-\theta)}\\
-\sqrt{\cal RT}e^{-i(\phi-\theta)}&-{\cal T} \end{array}\right),
\end{eqnarray}
where $\tau^3$ is the third Pauli matrix and
\begin{equation}
S=\left(\begin{array}{cc}r& t\\t & r'\end{array}\right)
=\left(\begin{array}{cc}\sqrt{\cal R}e^{i\theta}& \sqrt{\cal T}e^{i\phi}\\
\sqrt{\cal T}e^{i\phi} &-\sqrt{\cal R}e^{i(2\phi-\theta)} \end{array}\right)
\end{equation}
is the usual unitary scattering matrix in the presence of
time-reversal symmetry ($SS^\dagger = \hat{1}$ and $S=S^T$). In the
second equality the scattering probabilities ${\cal T}=|t|^2$ and
${\cal R}=1-{\cal T}=|r|^2=|r'|^2$ have been introduced. Eq.~(4) in
Ref.~\onlinecite{mort2001a} can now be written as

\begin{equation}
\Delta(\omega,x,x')=\pi^2\hbar\omega\frac{\hbar v_F}{L} {\rm Tr}\big(J \big[\rho(x);
\rho(x')\big]\big),
\end{equation}
where $[A;B]=AB-BA$ is a commutator with $A$ and $B$ being matrices. Next we employ a
unitary transformation
\begin{equation}
{\cal U}=\left(\begin{array}{cc}u & - v\\v^* & u^*\end{array}\right),
\end{equation}
with $|u|^2+|v|^2=1$ that satisfies ${\cal U}J {\cal U}^\dagger
=\sqrt{{\cal T}}\tau^3$ by choosing $|u|^2=\frac{1}{2}(1+\sqrt{\cal
  T})$, $|v|^2=\frac{1}{2}(1-\sqrt{\cal T})$, and
$vu^*=\frac{1}{2}\sqrt{\cal R}e^{i(\phi-\theta)}$ (in a concrete
calculation it can be useful to use the freedom to choose the phases
as $v=|v|e^{i(\phi-\theta)/2}$ and $u=|u|e^{-i(\phi-\theta)/2}$). It
is then easy to obtain the very compact result

\begin{equation}\label{Delta_tilde}
\Delta(\omega,x,x')=4\pi^2\hbar\omega\frac{\hbar v_F}{L}\sqrt{\cal T}i
{\rm Im}\big\{\tilde\rho_{+-}(x)\tilde\rho_{-+}(x')\big\},
\end{equation}
where in the new basis $\tilde\rho(x)={\cal U}^\dagger
\rho(x){\cal U}$. With the choice of relative phase mentioned
above we have in particular

\begin{eqnarray}\label{rho_tilde}
&&\tilde\rho_{+-}(x)= \frac{e^{i(\phi-\theta)}}{2}\Big[\sqrt{\cal R}
\big[\rho_{--}(x)-\rho_{++}(x)\big]\nonumber\\
&&\quad\quad+(1+\sqrt{\cal T})\rho_{+-}(x)-(1-\sqrt{\cal T})\rho_{+-}^*(x)\Big].
\end{eqnarray}

\widetext

\end{document}